\documentclass[twocolumn]{aastex63}
\usepackage{amssymb,amsmath}
\usepackage{graphicx}
\usepackage{natbib}
\usepackage[OT1,T1]{fontenc}
\usepackage{threeparttable}
\usepackage{multirow}

\begin{document}

\title{Revisiting the Spectral Features of Ellerman Bombs and UV Bursts.\\ I. Radiative Hydrodynamic Simulations}

\author{Jie Hong}
\affiliation{School of Astronomy and Space Science, Nanjing University, Nanjing 210023, People's Republic of China}
\affiliation{Key Laboratory for Modern Astronomy and Astrophysics (Nanjing University), Ministry of Education, Nanjing 210023, People's Republic of China}

\author{Ying Li}
\affiliation{Key Laboratory of Dark Matter and Space Astronomy, Purple Mountain Observatory, Chinese Academy of Sciences, Nanjing 210033, People's Republic of China}

\author{M.~D. Ding}
\affiliation{School of Astronomy and Space Science, Nanjing University, Nanjing 210023, People's Republic of China}
\affiliation{Key Laboratory for Modern Astronomy and Astrophysics (Nanjing University), Ministry of Education, Nanjing 210023, People's Republic of China}

\author{Qi Hao}
\affiliation{School of Astronomy and Space Science, Nanjing University, Nanjing 210023, People's Republic of China}
\affiliation{Key Laboratory for Modern Astronomy and Astrophysics (Nanjing University), Ministry of Education, Nanjing 210023, People's Republic of China}

\email{jiehong@nju.edu.cn}

\begin{abstract}
Ellerman bombs (EBs) and UV bursts are both small-scale solar activities that occur in active regions. They are now believed to form at different heights in the lower atmosphere. In this paper, we use one-dimensional radiative hydrodynamic simulations to calculate various line profiles in response to heating in different atmospheric layers. We confirm that heating in the upper photosphere to the lower chromosphere can generate spectral features of typical EBs, while heating in the mid to upper chromosphere can generate spectral features of typical UV bursts. The intensity evolution of the H$\alpha$ line wing in EBs shows a rise--plateau pattern, while that of the \ion{Si}{4} 1403 \AA\ line center in UV bursts shows a rise--fall pattern.
 However, the predicted enhancement of FUV continuum near 1400 \AA\ for EBs is rarely reported and requires further observations to check it. With two heating sources or an extended heating source in the atmosphere, both EB and UV burst features could be reproduced simultaneously.

\end{abstract}


\section{Introduction}
Solar active regions host the majority of  solar activities. Besides the large-scale energetic activities like solar flares and coronal mass ejections, many small-scale activities have also been observed. Ellerman bombs (EBs) are  one of them that were first observed by \cite{1917ellerman}, with typical features of enhanced H$\alpha$ line wings and an undisturbed line center. Subsequent observations showed that EBs are also visible in many other spectral lines or continua, including the line wings of the \ion{Ca}{2} 8542 \AA\ line \citep{2006fang,2013vissers,2013yang}, the \ion{Ca}{2} H and K lines \citep{2008matsumoto,2015rezaei}, the \ion{He}{1} D$_3$ and 10830 \AA\ lines \citep{2017libbrecht}, and the \ion{Mg}{2} triplet lines \citep{2015vissers,2017hansteen,2017honga}, as well as the G band \citep{2011herlender,2013nelson} and ultraviolet continua at 1600 \AA\ and 1700 \AA\citep{2013vissers,2015rezaei,2016tian,2017chen}. Yet EBs are {invisible} in neutral metal lines \citep{2015rutten} and coronal lines \citep{2013vissers,2016tian}.

Since the visibility of EBs varies in different wavelengths, one can deduce the origin height of EBs from various spectral lines with different formation heights. \cite{1917ellerman} suggested that EBs must occur somewhere lower than the reversing layer of the solar atmosphere, since the H$\alpha$ line center is not disturbed. Non-local thermodynamic equilibrium models indicated that in order to reproduce the observed EB line profile, there must be heating in the lower atmosphere (near the temperature minimum region (TMR)), and the local temperature enhancement is around 600--3000 K \citep{2006fang,2014berlicki}. Recent 1D radiative hydrodynamic (RHD) models obtained similar results \citep{2017reid,2017hong}. The total energy of EBs are calculated to be in a range of $10^{22}$--$10^{28}$ erg \citep{2002georgoulis,2006fang,2013nelsona,2016reid}.

It seems that a consensus has been reached about the EB models when the first report of ``hot explosions'' came into sight \citep{2014peter}. These ``hot explosions'', now more commonly referred to as UV bursts, have very strong and wide \ion{Si}{4} line profiles with superposed absorptive metal lines (\ion{Ni}{2} and \ion{Fe}{2}). To explain these spectral features, \cite{2014peter} proposed that these activities are rooted in the deep atmosphere, and the expanding hot plasma due to magnetic reconnection gives rise to the \ion{Si}{4} lines, while the less disturbed chromospheric plasma causes the metal absorption lines. Further observations showed that some UV bursts and EBs have co-spatial and simultaneous appearance \citep{2015kim,2015vissers,2016tian}, which has led to a direct impression that EBs and UV bursts are possibly the same activities. This would propose a big challenge to the traditional EB model, since the response in the \ion{Si}{4} line  requires a local temperature of at least the order of $10^4$ K. Although magnetohydrodynamic (MHD) simulations have proved that even in the partially-ionized photosphere it is still possible to reach such a high temperature \citep{2016ni}, there were still voices against it, claiming that such a high temperature below the chromosphere would destroy the typical features of an EB in the H$\alpha$ line wing \citep{2017fang,2017hong}. 1D RHD simulations also showed that it is impossible to reproduce the spectral features of EBs and UV bursts with only one heating layer in the lower atmosphere \citep{2017reid}.

More recent 3D RMHD simulations suggested that EBs and UV bursts are actually formed at different heights, and the formation height of the UV bursts is rendered to be within the mid to upper chromosphere \citep{2017hansteen,2019hansteen}. The synthetic line profiles in those simulations have reproduced some typical features of the observed ones, including the \ion{Ni}{2} absorption line, which is explained  as a contribution from the slow-rising cool dense gas between the heated chromospheric layer and the transition region \citep{2019hansteen}. Near-limb observations also suggest that UV bursts are formed higher than EBs \citep{2019chen}. The observation  of a UV burst without any EB features showed that the H$\alpha$ line center can be influenced by a surge related to the UV burst \citep{2020ortiz} .

If EBs and UV bursts are not formed at the same height, then it would be very important to distinguish the spectral features of EBs and UV bursts {to avoid any misinterpretation}, especially when they are observed co-spatially. Apart from realistic 3D simulations, 1D RHD simulations still prove to be an effective way to explore the spectral features of these events, since it is feasible to manually control the energy input. This is a  reasonable simplification if one is only concerned about the results of the energy deposition, but not the detailed physical processes. In this paper, we use the 1D RHD models to calculate the line profiles in response to various heating regions in the atmosphere. The line profiles are then compared with previously observed ones of EBs and UV bursts, in order to better constrain their formation heights. The rest of the paper is organized as follows. In Section 2, we briefly introduce our method. Then we show our main results in Section 3. A discussion of our results compared with other simulations and observations is given in Section 4, followed by a final conclusion in Section 5.

\section{Method}
\begin{figure}
\epsscale{1.2}
\plotone{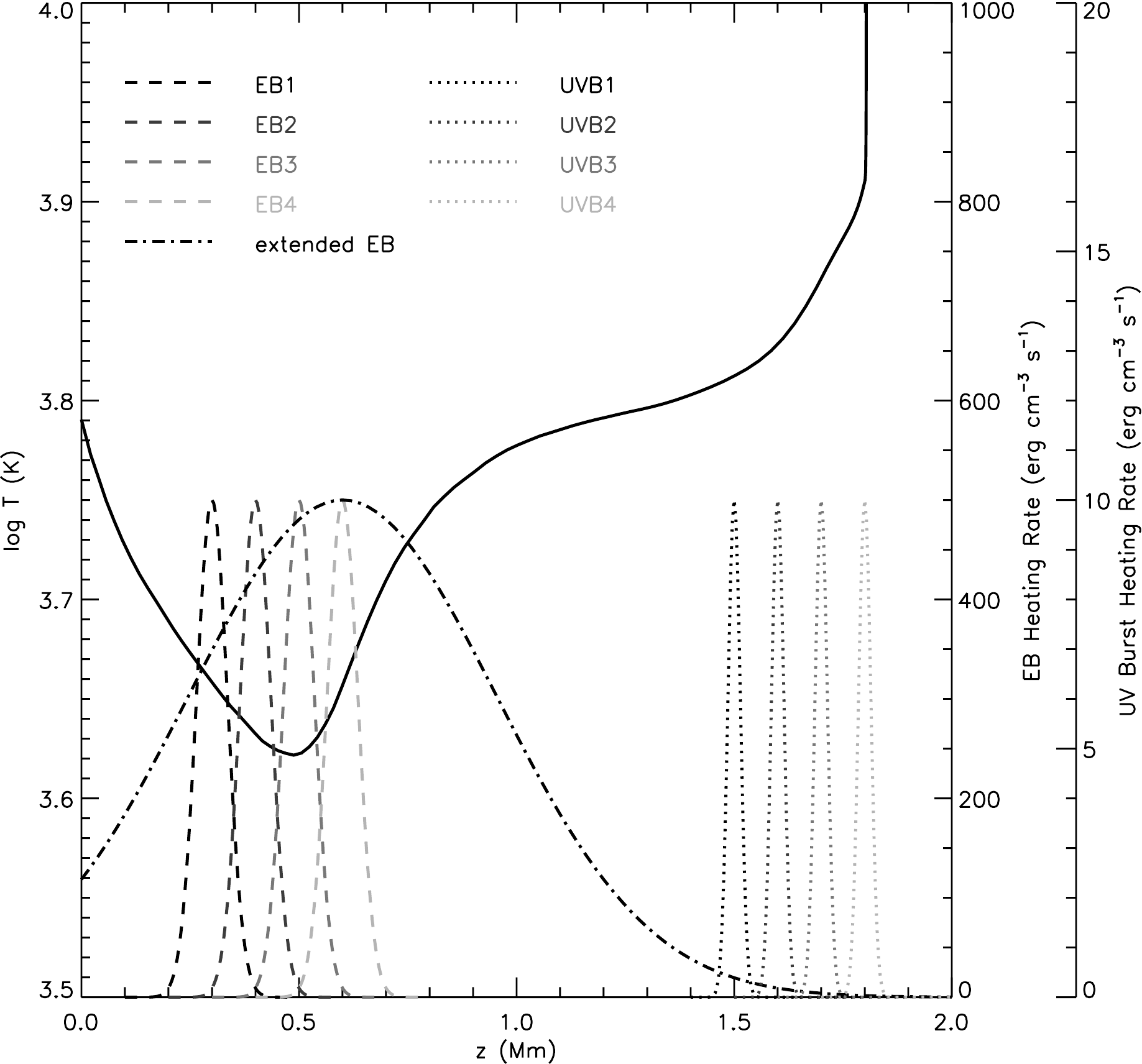}
\caption{The height distribution of temperature at $t=0$ s (solid), as well as the heating rate in each EB (dashed and dash-dotted) and UV burst (dotted) case.  }
\label{heatingrate}
\end{figure}

We use the 1D RHD code \verb"RADYN" \citep{1992carlsson,1995carlsson,1997carlsson,2002carlsson} to model EBs and UV bursts. \verb"RADYN" can solve the hydrodynamic equations, the population rate equation and the radiative transfer equation together implicitly with an adaptive grid, in a 10 Mm quarter-circular loop. The initial atmosphere is the VAL-based quiet-Sun model \citep{2017hong}, with a loop-top temperature of 1 MK. Following \cite{2017hong}, the heating of EBs and UV bursts is modeled with an additional term in the energy equation that is height-dependent. We assume that the heating rate is constant and has a Gaussian shape along height. The peak of the heating rate ($Q_\mathrm{peak}$) is set to 500 erg cm$^{-3}$ s$^{-1}$ for EBs and 10 erg cm$^{-3}$ s$^{-1}$ for UV bursts. The width of the heating profile {(defined as the half width at 1/e maximum)} is set to 50 km for EBs and 20 km for UV bursts. The center of the heating profile varies from 300 km to 600 km for EBs, which ranges from the upper photosphere to the lower chromosphere; while it varies from 1.5 Mm to 1.8 Mm for UV bursts, which ranges from the upper chromosphere to the transition region.  Figure~\ref{heatingrate}  plots the height distribution of the heating rate in each case with the initial temperature profile. The heating rate of EBs is similar to \cite{2017reid} and \cite{2017hong}. We discuss our choice of the heating parameters further in Section~\ref{param}. The excitation and ionization from non-thermal electrons are included in the calculation following \cite{1993fang}. All the models are run for 10 s, and we save the snapshots every 0.1 s. Then we select four lines that are frequently used in observations, namely the H$\alpha$ line, the \ion{Si}{4} 1403 \AA\ line, the \ion{Mg}{2} 2791 \AA\ line, and the \ion{Mg}{2} k line, and calculate their responses in detail.

Normally, both the non-equilibrium ionization (NEI) and partial frequency redistribution (PRD) effects can influence the calculated line profiles. \verb"RADYN" has included in the calculation the NEI effect while assumes a Gaussian profile for the Lyman series to mimic the PRD effect. This turns out to be a reasonable approximation \citep{2012leenaarts} and we take the H$\alpha$ line profiles directly from \verb"RADYN". 

For the \ion{Si}{4} lines that are not included in the main run of \verb"RADYN", we use \verb"MS_RADYN" to rerun the whole simulation with the silicon atom, where we follow the main run step by step and solve the radiative transfer of the new atom \citep{2019kerrb}. The results from \verb"MS_RADYN" are then imported to \verb"RH" \citep{2001uitenbroek,2015pereira} for a better treatment of the \ion{Si}{1} continuum near the \ion{Si}{4} lines. Similar to \cite{2019hong}, in \verb"RH" we stop to solve the statistical equilibrium equation for Si, in order to keep the NEI effect that is present in \verb"MS_RADYN". 

For the \ion{Mg}{2} lines that are largely affected by PRD, we again use \verb"RH". Here we need to solve the statistical equilibrium equation for Mg, since the level populations are not known. The NEI effects for this atom are thus absent but we try to mitigate this by including the hydrogen and electron density from \verb"RADYN" \citep{2019kerr,2020hong}.

\begin{figure}
\epsscale{1.2}
\plotone{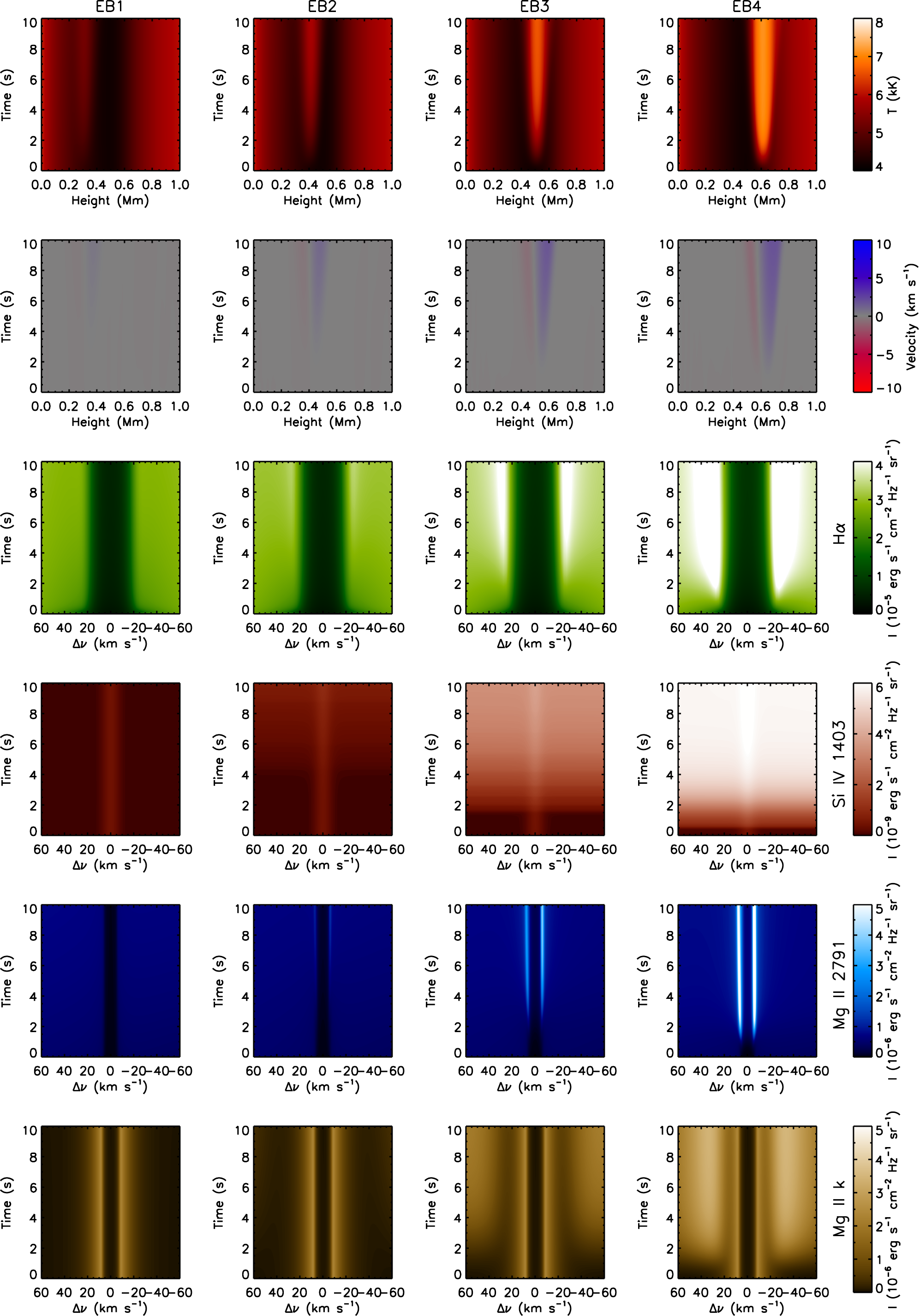}
\caption{Time evolution of the height distributions of temperature and velocity for the EB cases, as well as the emergent line intensities. Positive velocity values denote upward motions, while negative velocity values denote downward motions. The horizontal axes of the line profile panels are in Doppler scale.}
\label{evolebatm}
\end{figure}

\begin{figure}
\epsscale{1.2}
\plotone{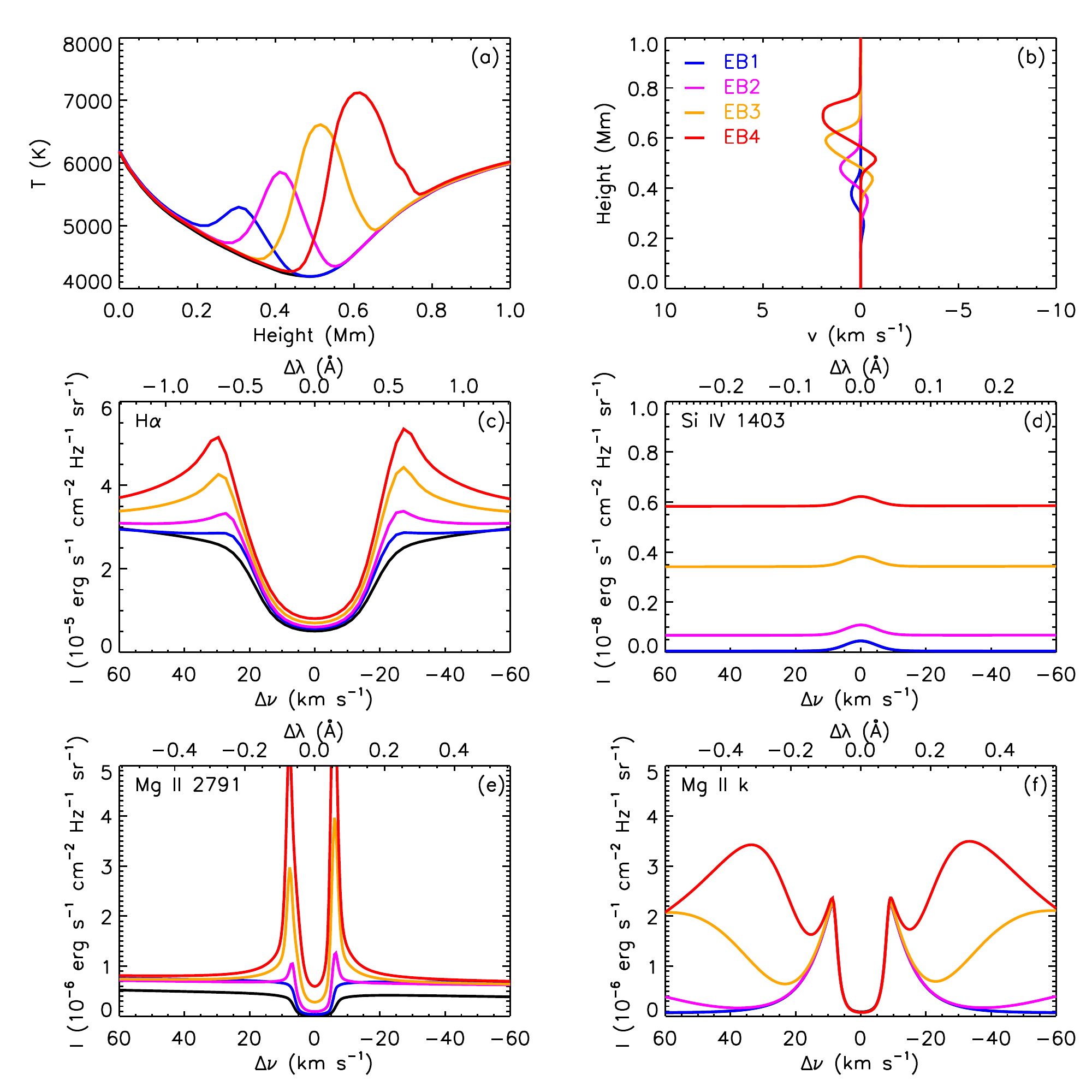}
\caption{Height distributions of temperature and velocity, as well as the synthetic line profiles, for the EB cases at $t=0$ s (black) and $t=10$ s (colored). The sign of velocity in panel (b) has the same definition as in Fig.~\ref{evolebatm}. The bottom horizontal axis in panels (c)--(f) is in Doppler scale. {Note that the blue line is on top of the black line, and in panels (d) and (f) they appear to be overlapped.}}
\label{eball}
\end{figure}

\begin{figure}
\epsscale{1.2}
\plotone{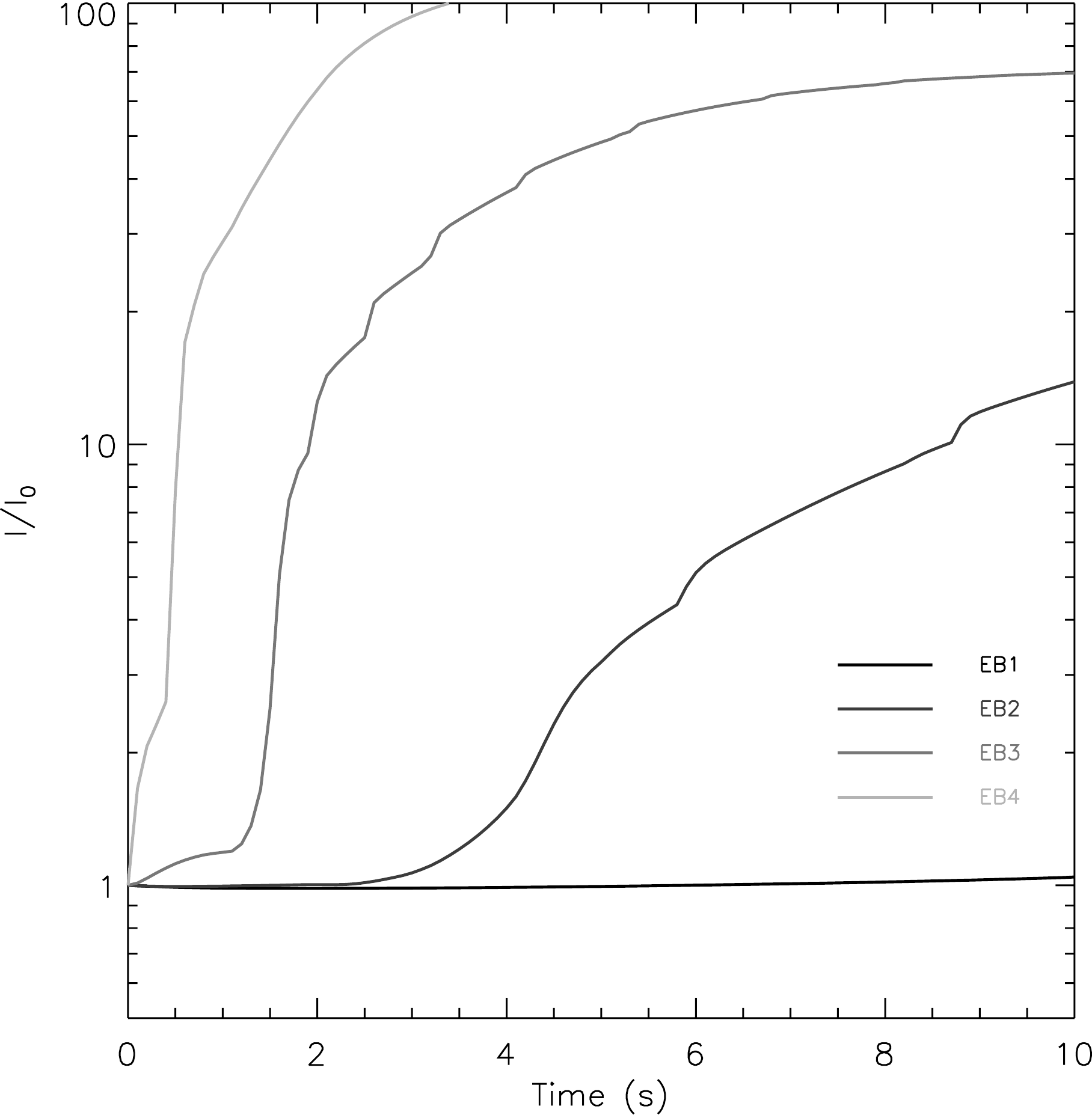}
\caption{Time evolution of the FUV continuum intensity near the \ion{Si}{4} 1403 \AA\ line for the EB cases.}
\label{si4cont}
\end{figure}

\begin{figure}
\plotone{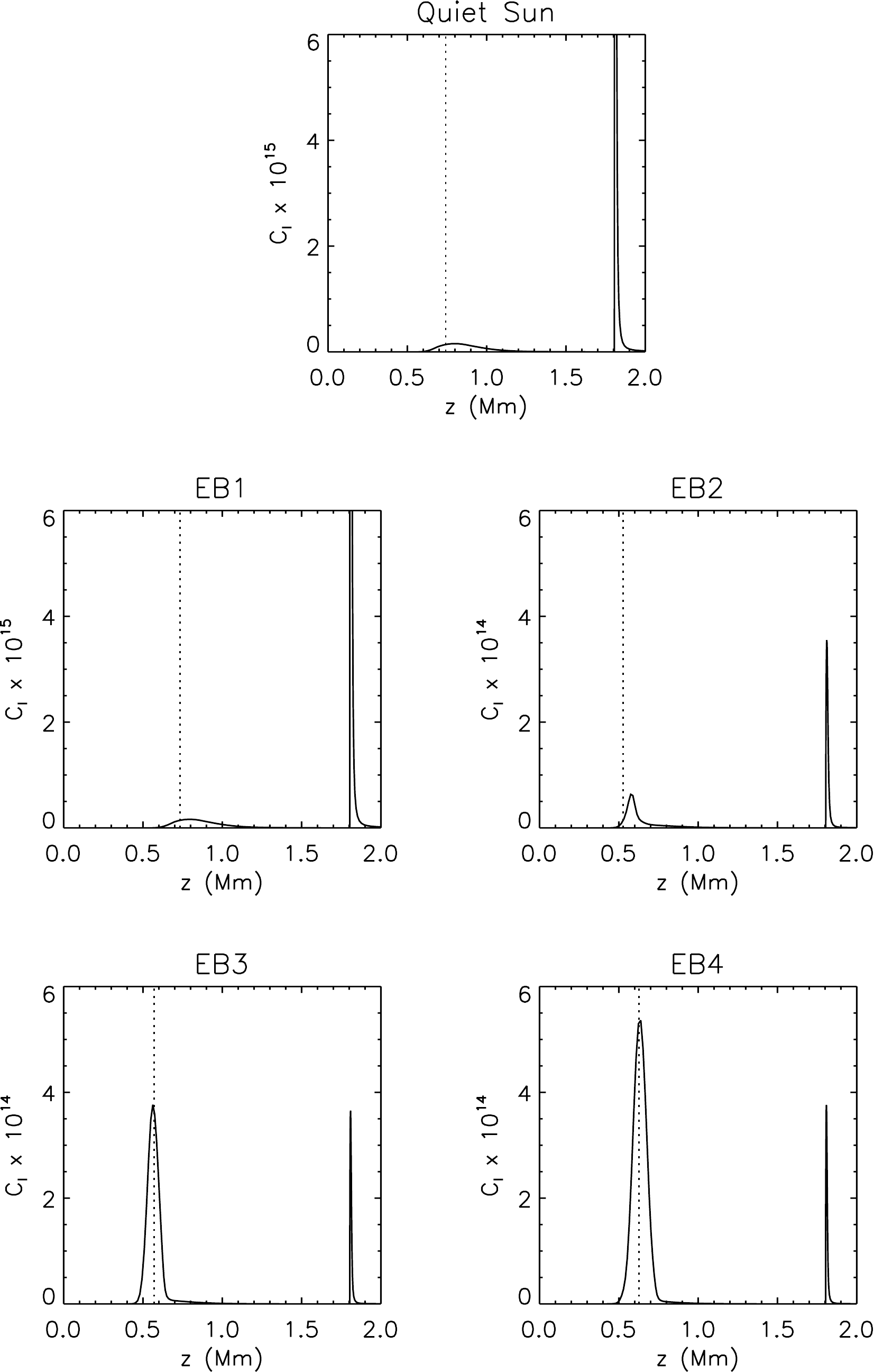}
\caption{Contribution function of the \ion{Si}{4} 1403 \AA\ line center  in the quiet Sun and for the EB cases at $t=10$ s. The vertical dotted lines denote the height where $\tau=1$.}
\label{si4form}
\end{figure}

\section{Results}
\label{res}
\subsection{The EB cases}
In Fig.~\ref{evolebatm} we show the time evolution of the height distributions of temperature and velocity for the EB cases (see the top two panels). As expected, local heating increases the temperature, and the resulted pressure gradient pushes plasma both upwards and downwards, forming bidirectional mass flows. The magnitude of heating in the four cases is the same, while the temperature enhancement and the velocity are different from case to case, which is the largest in EB4, but the smallest in EB1. At the end of heating, the temperature of EB1 reaches 5300 K (with a temperature enhancement $\Delta T=700$ K), while the temperature of EB4 exceeds 7000 K (with a temperature enhancement $\Delta T=2400$ K) (Fig.~\ref{eball}). This is because EB4 is formed in the lower chromosphere, whose density is smaller than that of the upper photosphere where EB1 is formed. The temperature enhancements in our EB cases are within the range of the classic EB models \citep{2006fang,2014berlicki}. The mass flow velocity, in general, is very small and does not exceed 2 km s$^{-1}$.

The time evolution of the line intensities is also shown in Fig.~\ref{evolebatm}. Specifically, we plot the line profiles at 0 s and 10 s in Fig.~\ref{eball}. Generally speaking, for EB2 to EB4, one can see clear  enhancements in the line intensities when time proceeds. For EB1, however, the enhancements are very faint. 

\begin{figure}
\plotone{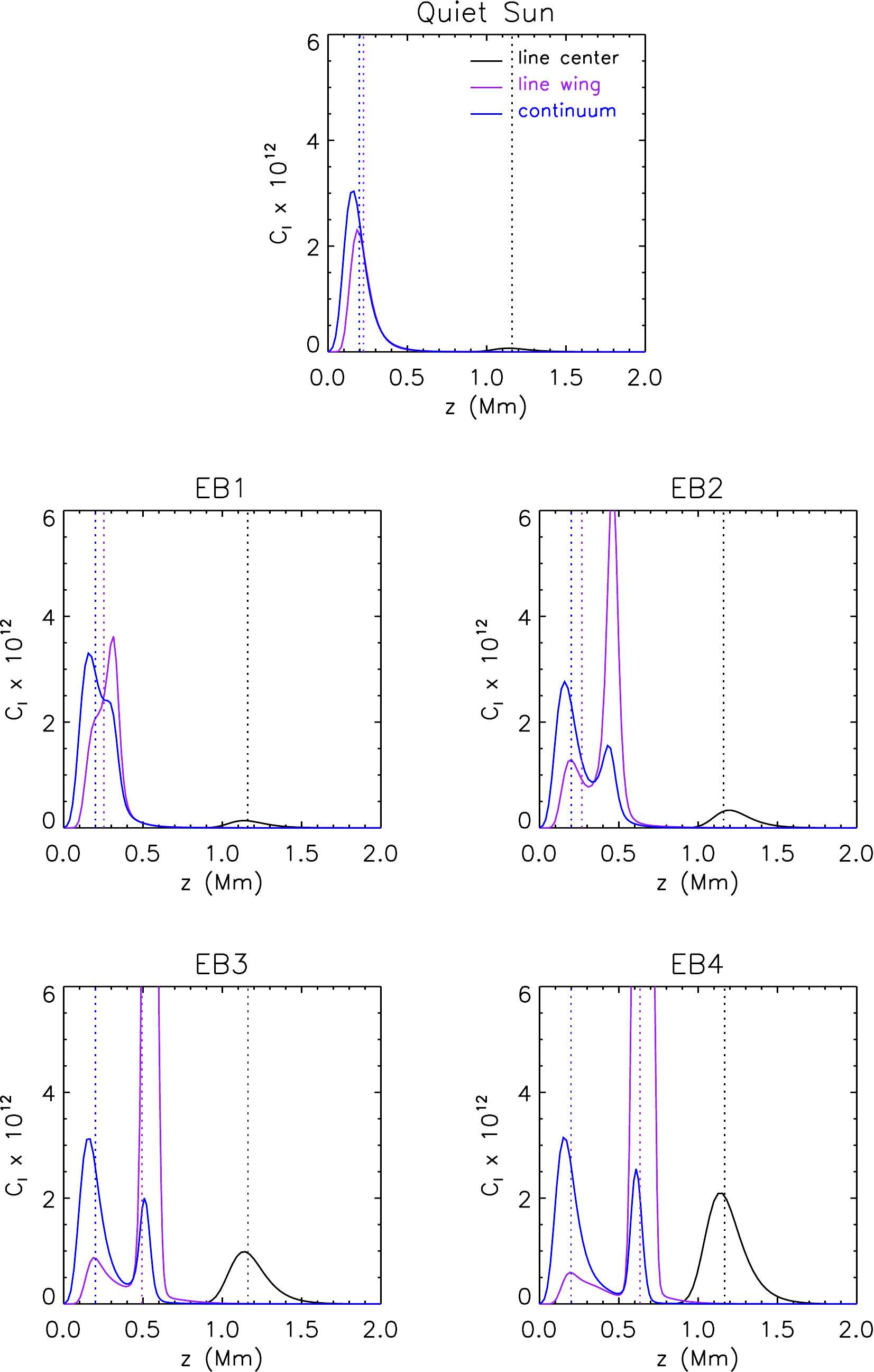}
\caption{Contribution function of the \ion{Mg}{2} 2791 \AA\ line center (black), the near wing ($-0.075$ \AA, purple) and the nearby NUV continuum ($-0.56$  \AA, blue) in the quiet Sun and for the EB cases at $t=10$ s. The vertical dotted lines denote the height where $\tau=1$. }
\label{mg2form}
\end{figure}

All four EB cases present an enhancement in the H$\alpha$ line wing (Fig.~\ref{evolebatm} and Fig.~\ref{eball}(c)), which is a typical observational feature of EBs. However, the magnitude of  line wing enhancement varies with cases, the relative increase is 20\% for EB1, and can reach 110\% for EB4. For EB2 to EB4, the line wings show two obvious peaks, while for EB1,   the shape of the line wing is flat without any obvious peak. The H$\alpha$ line center   is also enhanced in these EB cases, with a relative increase of 10\% for EB1, and 60\% for EB4. The increase in the H$\alpha$ line core intensity is not considered as an EB feature, and most observations show undisturbed H$\alpha$ line core. However, there are indeed some observations that report a relative increase of up to 60\% in the H$\alpha$ line center \citep{2013vissers,2016grubecka}.

The most evident feature in the response of the \ion{Si}{4} line is the enhancement of the whole line including the FUV continuum (Fig.~\ref{evolebatm} and Fig.~\ref{eball}(d)). In Fig.~\ref{si4cont} we show the time evolution of the FUV continuum intensity. For EB1, there is very little enhancement, while for EB2 to EB4, a sharp increase is present. The intensity can rise by more than two orders of magnitude in EB4.

In Fig.~\ref{si4form} we plot the contribution function at the \ion{Si}{4} 1403 \AA\ line center. The intensity of the \ion{Si}{4} line actually consists of radiation from both the FUV continuum and the optically thin spectral line.  The FUV continuum, mainly contributed by \ion{Si}{1} bound-free transitions, is formed in the lower chromosphere. Thus heating in the lower chromosphere (EB2 to EB4) can contribute to the FUV continuum, while heating in the photosphere (EB1) has little contribution. The optically thin line, however, is formed in the transition region, where EB heating has little influence. Therefore, the enhancement of the \ion{Si}{4} 1403 \AA\ line here is merely a result of the FUV continuum enhancement.

The \ion{Mg}{2} 2791 \AA\ line shows enhancement in the line wing and the nearby NUV continuum (Fig.~\ref{evolebatm} and Fig.~\ref{eball}(e)). Similar to the response of the H$\alpha$ line, the line wing of the \ion{Mg}{2} 2791 \AA\ line is flat for EB1, while two obvious peaks exist for EB2 to EB4. The two peaks display a weak asymmetry that originates from mass flows driven by the pressure gradient. In addition, the line center is also enhanced. 

The contribution functions of the \ion{Mg}{2} 2791 \AA\ line center, the near wing and the nearby NUV continuum are plotted in Fig.~\ref{mg2form}. In the pre-EB atmosphere, the line center is formed in the mid chromosphere at around 1.2 Mm, and the near wing and the NUV continuum are formed in the photosphere. When heating sets in, the opacity is increased, and the $\tau=1$ height of the near wing is levitated. The heated region now has a large contribution to the near wing intensity and the NUV continuum, and the contribution function  shows a second peak near the heated region. The spectral response of this line resembles that of the H$\alpha$ line very much.

The \ion{Mg}{2} k$_3$ {core} and k$_2$ peaks are formed in the mid to upper chromosphere, and thus they do not show any response to EB heating. The outer wings, however, can be enhanced since they form lower than the k$_2$ peaks (Fig.~\ref{evolebatm} and Fig.~\ref{eball}(f)). Therefore, deep photospheric heating (EB1) has little contribution to the line wing intensity. However, when the heating region moves higher (from EB2 to EB4), one can notice a gradual enhancement in  the wings of this line.

\begin{figure}
\epsscale{1.2}
\plotone{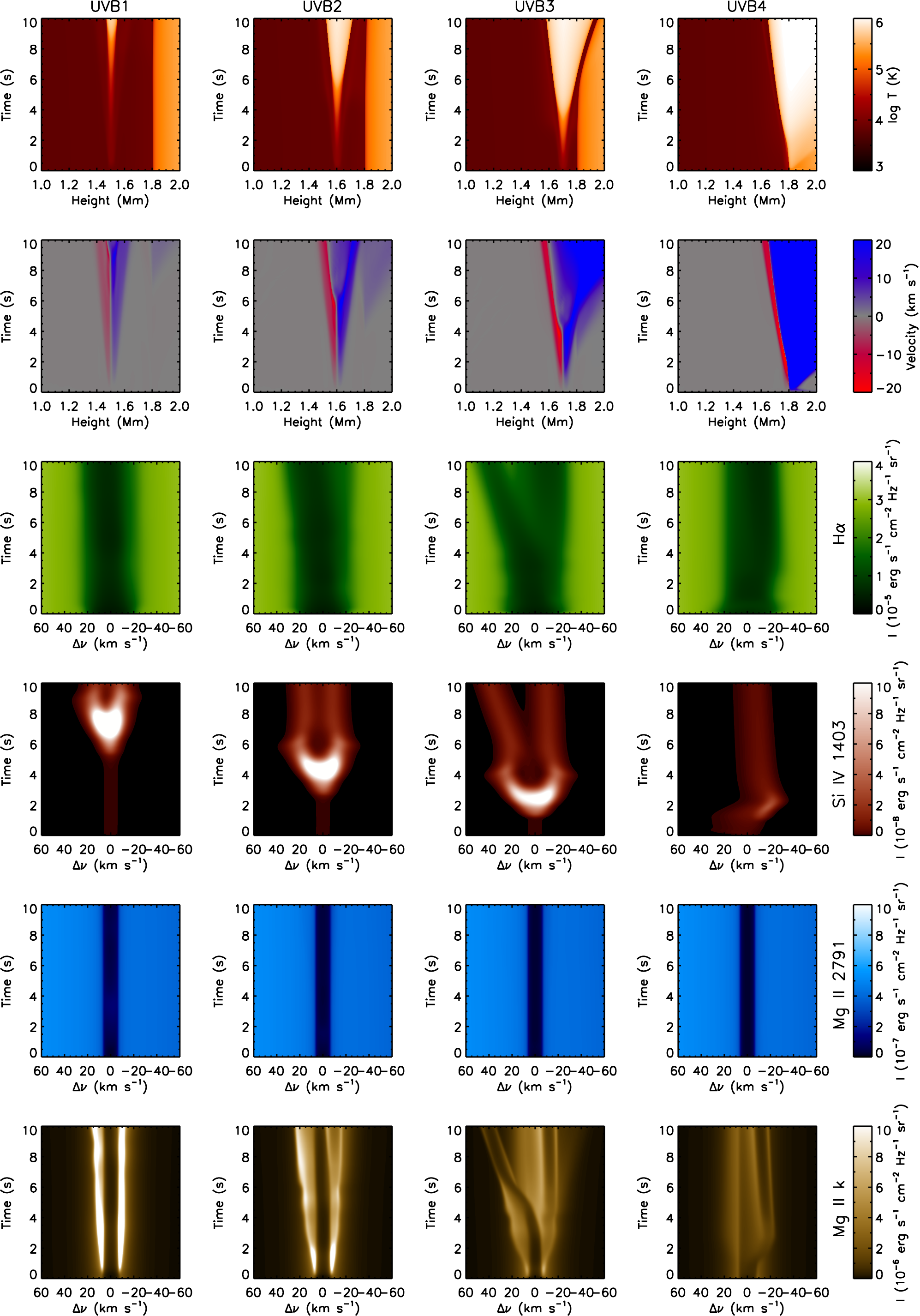}
\caption{Same as Fig.~\ref{evolebatm}, but for the UV burst cases.}
\label{evolibatm}
\end{figure}

\begin{figure}
\epsscale{1.2}
\plotone{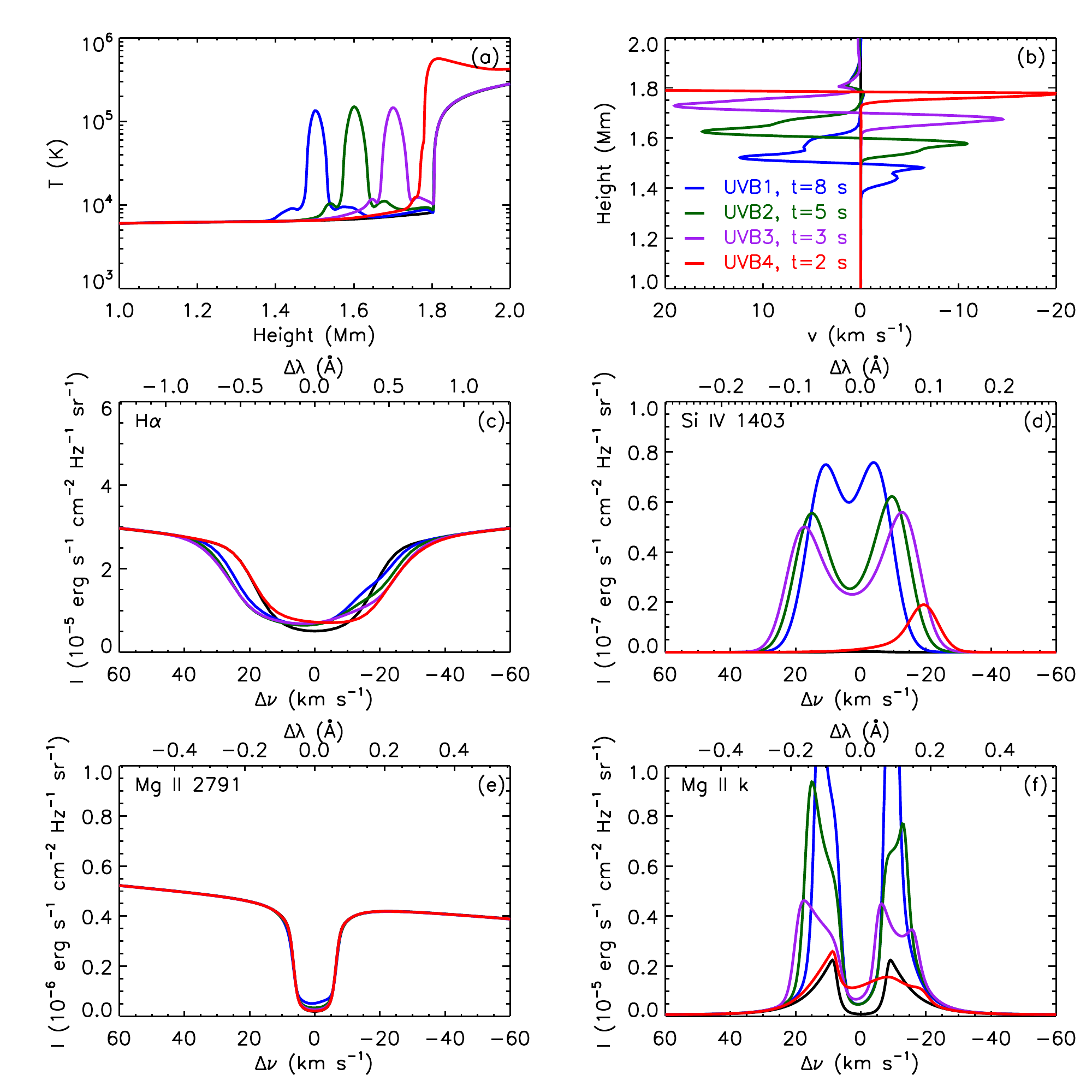}
\caption{Same as Fig.~\ref{eball}, but for the UV burst cases. Note that the selected time for plotting the line profiles is different in the four cases (colored lines).}
\label{iball}
\end{figure}

\begin{figure}
\plotone{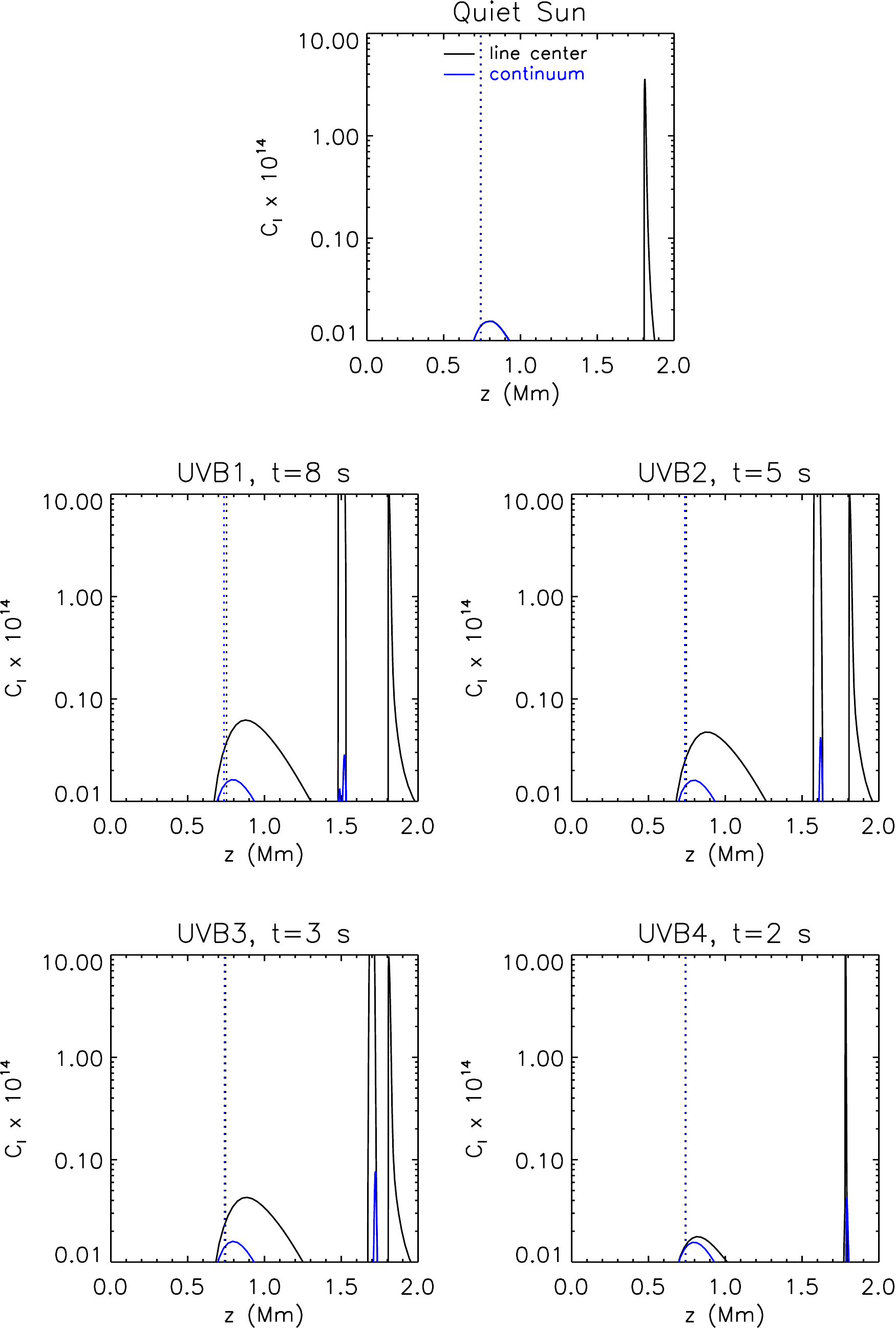}
\caption{Contribution function of the \ion{Si}{4} 1403 \AA\ line center (black) and the nearby FUV continuum ($- 0.2$ \AA, blue)  in the quiet Sun and for the UV burst cases at $t=10$ s. Note that the selected time for plotting the contribution functions is different in the four cases. The vertical axes are in logarithmic scale.}
\label{si4form2}
\end{figure}

\subsection{The UV burst cases}
We show the time evolution of temperature, velocity and line profiles of the UV burst cases in Fig.~\ref{evolibatm}. Heating of UVB1 to UVB3 is located in the upper chromosphere, while in the transition region for UVB4. As expected, the temperature of the heated region gets enhanced, and the pressure gradient drives bidirectional mass flows. The temperature can exceed 10$^5$ K and approach 10$^6$ K in some cases, and the flow velocity can reach nearly 20 km s$^{-1}$.

The \ion{Si}{4} 1403 \AA\ line has been used to identify UV bursts. In our simulation, after a certain time of heating, this line indeed shows an enhanced and broadened profile (Fig.~\ref{evolibatm}). The local temperature of the heated region is around 10$^5$ K when the line intensity is at its maximum. As heating proceeds, the local temperature exceeds the formation temperature of this line, and thus the line intensity begins to decrease. 

In Fig.~\ref{iball} we show the line profiles of the four UVB cases at a specific time when the \ion{Si}{4} line is the strongest in each case. UVB1 to UVB3 all show a double-peaked line profile, as a result of the bi-directional flows in the chromosphere. UVB4, however, shows a single redshifted peak, corresponding to the downflow below the transition region. The hot upflow plasma is in the corona that has no contribution to this line. The nearby FUV continuum is also slightly enhanced, but not as evident as in the EB cases.

The contribution functions of the \ion{Si}{4} line and nearby continuum are plotted in Fig.~\ref{si4form2}. The FUV continuum is mostly formed in the lower chromosphere. However, heating in the upper chromosphere can also have some contribution, which may result in the continuum increase. The line center intensity of UVB1 to UVB3, is contributed by three parts. The first part is from the lower-mid chromosphere around 0.9 Mm that provides the continuum photons, and the second part is from the transition region that provides the \ion{Si}{4} line photons. The third part, which is also the dominant part, is from the heated chromosphere where most \ion{Si}{4} line photons originate. For UVB4 the latter two parts are merged into one. This line is optically thin in all the UVB cases, since most line photons originate in specific layers far above the $\tau=1$ height.

The \ion{Mg}{2} k line shows large enhancements of the k$_2$ peaks as well as k$_3$ (Fig.~\ref{evolibatm} and Fig.~\ref{iball}(f)). The k$_2$ peaks are not symmetric and k$_3$ is slightly blueshifted due to the velocity gradient in the chromosphere. The width of this line  also increases. However, heating does not influence the outer wings ($\Delta\nu>\pm30$ km s$^{-1}$). By comparison, the \ion{Mg}{2} 2791 \AA\ line wing and the nearby FUV continuum show no response in the four cases, although the line center can be slightly enhanced (Fig.~\ref{evolibatm} and Fig.~\ref{iball}(e)).

We also find that the H$\alpha$ line wing shows no response, but the line center is enhanced in the four cases (Fig.~\ref{evolibatm} and Fig.~\ref{iball}(c)). For UVB1 to UVB3, the line center is blueshifted, while in UVB4, the line center is redshifted.

\begin{table*}[ht]
\centering
\caption{Summary of spectral features in EBs and UV bursts in observations and simulations}
\begin{tabular}{c|c|c|c|c|c|c|c|c|c}
\hline
  &\multirow{2}{*}{Activity}& \multicolumn{2} {c|} { H$\alpha$ } & \multicolumn{2} {c|} {\ion{Mg}{2} triplet}    & \ion{Mg}{2}  & NUV  & \multirow{2}{*}{\ion{Si}{4}} & FUV   \\

 &  & wing & center & wing & center& k$_2$/h$_2$ peaks & continuum &  & continuum \\
\hline
\multicolumn{10} {c} { Observations } \\
\hline
\cite{2017hansteen} &\multirow{3}{*}{EB} & \checkmark  & $\times$ & \checkmark  &   & \checkmark & \checkmark & \checkmark & $\times$ \\
\cite{2017honga} & & \checkmark  & $\times$ & \checkmark  & \checkmark & $\times$ & \checkmark & $\times$ & $\times$ \\
\cite{2020ortiz} &  & \checkmark  & $\times$ & $\times$  & $\times$ & $\times$ & $\times$ & $\times$ & $\times$ \\
\hline
\cite{2014peter} &\multirow{4}{*}{UV burst}&    &  &  &  & $\checkmark$ &  & \checkmark &  \\
\cite{2016tian} & & $\times$  & $\times$ & \checkmark  & $\times$ & \checkmark & $\times$ & \checkmark & $\times$ \\
\cite{2017hansteen} &  & $\times$  & $\times$ & \checkmark  &   & \checkmark & \checkmark & \checkmark & $\times$ \\
\cite{2020ortiz} &  & $\times$  & \checkmark & $\times$  & $\times$ & \checkmark & $\times$ & \checkmark & $\times$ \\
\hline
\cite{2015vissers} &\multirow{4}{*}{Both}& \checkmark  & $\times$ & \checkmark  &  & \checkmark & \checkmark & \checkmark & $\times$ \\
\cite{2016tian}  & & \checkmark  & $\times$ & \checkmark  & \checkmark & \checkmark & \checkmark & \checkmark & \checkmark \\
\cite{2019chen} &   & \checkmark  & $\times$ & \checkmark  &   & \checkmark & \checkmark & \checkmark & \checkmark \\
\cite{2020ortiz} &   & \checkmark  & $\times$ & \checkmark  &   & \checkmark & \checkmark & \checkmark & $\times$ \\
\hline
\multicolumn{10} {c} { Simulations } \\
\hline
\cite{2017hansteen} &\multirow{5}{*}{EB} & \checkmark  & $\times$ & \checkmark  &  \checkmark & \checkmark & \checkmark & \checkmark$^\textrm{a}$ & \checkmark$^\textrm{a}$ \\

\cite{2017reid} &  & \checkmark  & \checkmark &    &    & \checkmark & \checkmark &  &  \\
\cite{2017hong} &  & \checkmark  & \checkmark &    &    &   &   &   &  \\
\cite{2019hansteen} &  & \checkmark  &   &    &    &   &   &   &  \\
\cite{2019seo} &  & \checkmark  & $\times$ &    &    &   &   &   &  \\
This paper &  & \checkmark  & \checkmark & \checkmark   &  \checkmark  &  $\times$ & \checkmark  &  \checkmark & \checkmark \\
\hline
\cite{2017hansteen} &\multirow{4}{*}{UV burst} & $\times$  & \checkmark & \checkmark  &   & \checkmark & \checkmark & \checkmark & $\times$ \\
\cite{2019hansteen} &    &   &  &   &   &  &  &\checkmark  & $\times$ \\
\cite{2021ni} &    &   &  &   &   &  &  &\checkmark  &  \\
This paper &    & $\times$  &\checkmark  & $\times$  &  \checkmark & \checkmark &$\times$  &\checkmark  & $\times$ \\
\hline
This paper &  Both  & \checkmark  &\checkmark  & \checkmark  &  \checkmark & \checkmark &\checkmark  &\checkmark  & \checkmark \\
\hline
\end{tabular}
\begin{tablenotes}
\item {Note.} Early observations  without  IRIS  are not included in this table.  The ``\checkmark'' sign indicates that there is obvious enhancement of the intensity, and the ``$\times$'' sign means that there is hardly any response. It is left blank if the wavelength is not used or it is hard to judge if there is a definite enhancement.
\item $^\textrm{a}$ By heating events unrelated to EBs.
\end{tablenotes}
\label{table2}
\end{table*}

\section{Discussion}
In Table~\ref{table2} we list the spectral features of EBs and UV bursts from our results as well as previous observations and simulations. As one can see, our simulations agree with previous studies generally, with a few exceptions. These agreements and exceptions are discussed below.

\begin{figure}
\epsscale{1.2}
\plotone{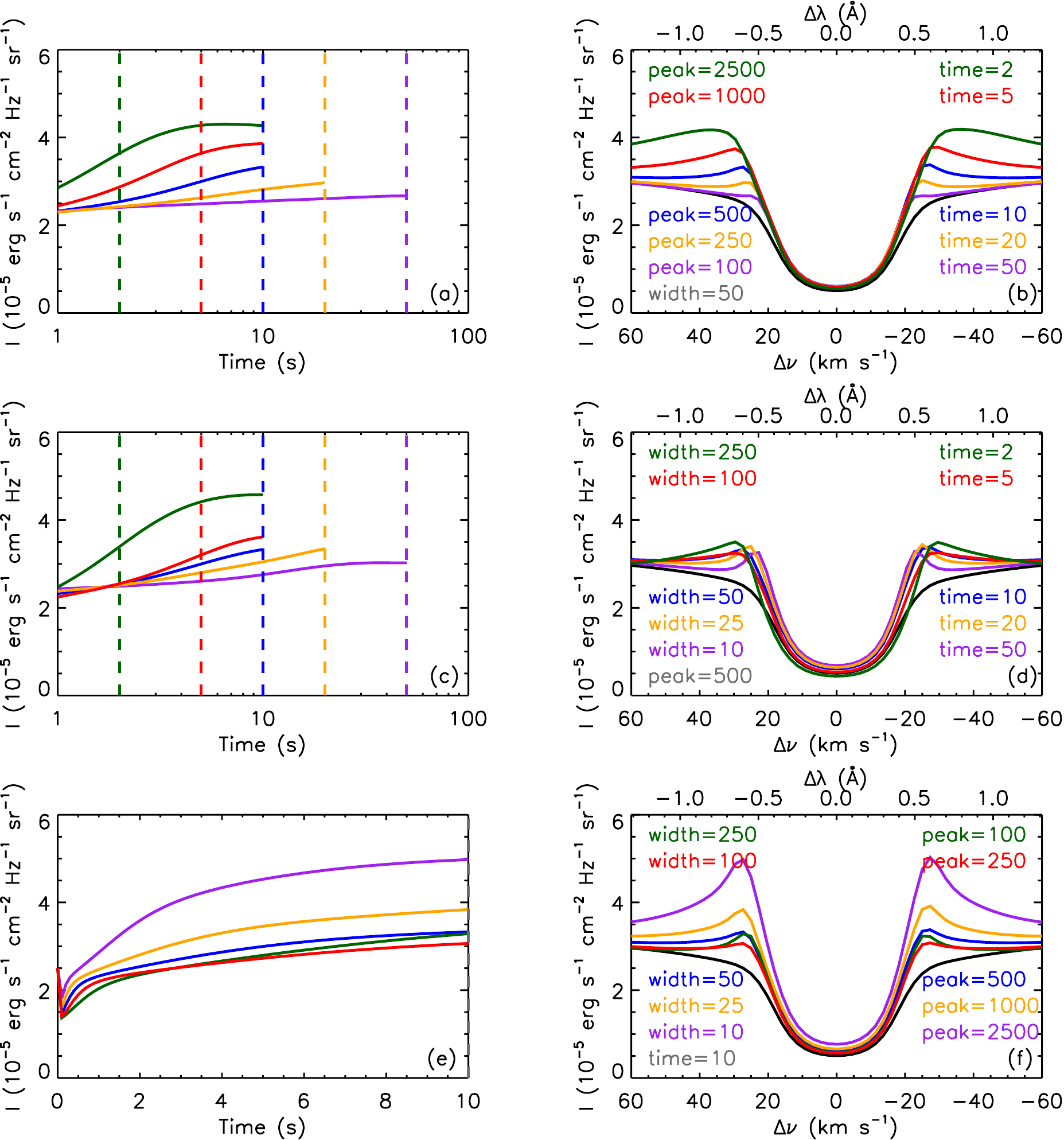}
\caption{Comparison of  the H$\alpha$ line wing ($\Delta\nu=30$ km s$^{-1}$) intensity (left column) and line profiles (right column) for different cases with varying heating parameters. The blue lines denote the case of EB2 as a reference, and the black lines denote the line profiles for the quiet Sun. The peak and width of the heating profile change from 20\% to 500\% of the values of case EB2, while the center of the heating profile remains to be 0.4 Mm. The total energy input for each of the line profiles shown in the right column is identical. The time for each line profile in the right column is marked with vertical dashed lines in the left column. }
\label{eb02all}
\end{figure}

\begin{figure}
\epsscale{1.2}
\plotone{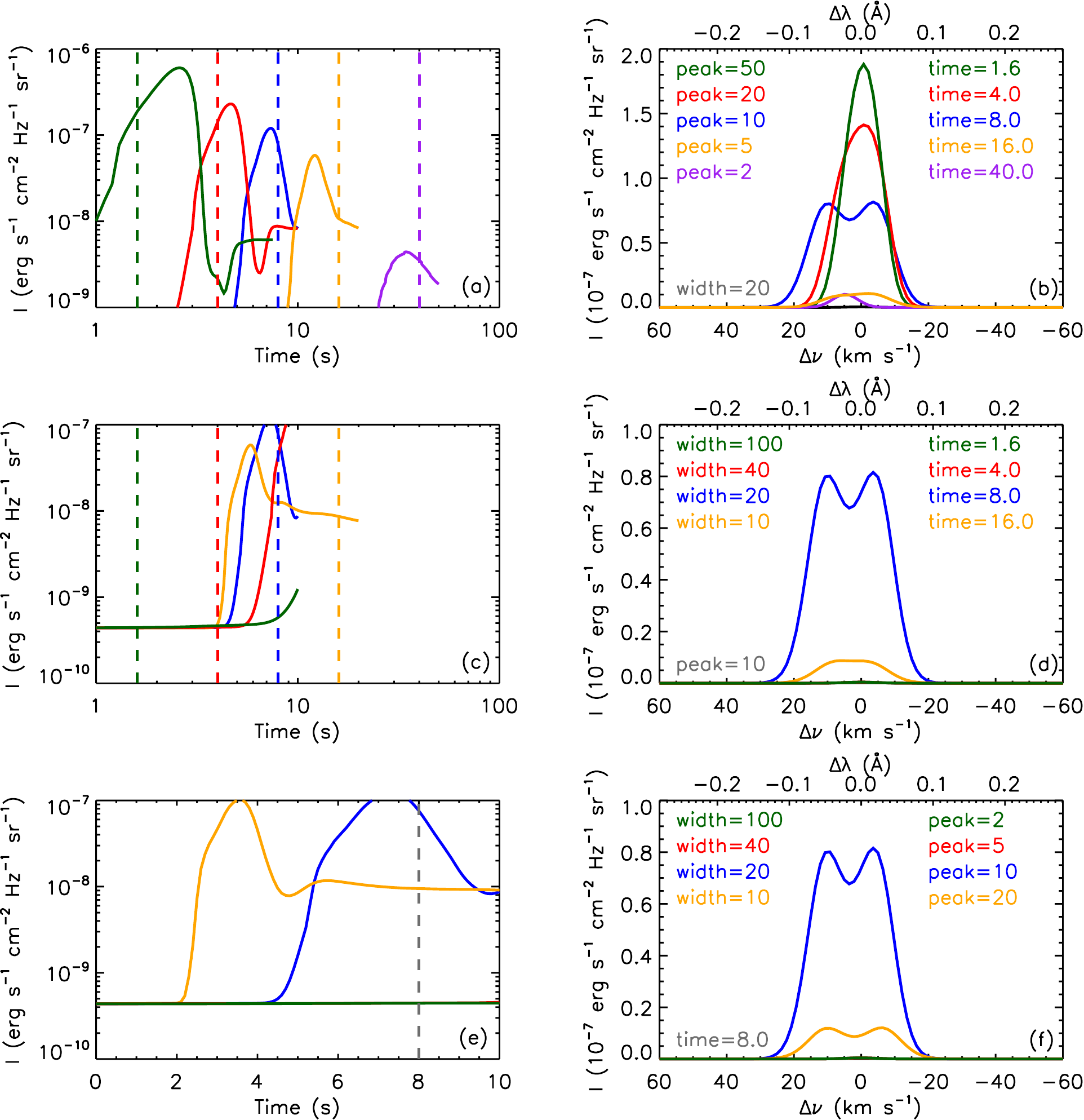}
\caption{Comparison of  the \ion{Si}{4} 1403 \AA\ line center intensity (left column) and line profiles (right column) for different cases with varying heating parameters. The blue lines denote the case of UVB1 as a reference. The quiet-Sun line profiles are too weak to be seen in these panels. The peak of the heating profile changes from 20\% to 500\%, and the width of the heating profile changes from 50\% to 500\% of the values of case UVB1, while the center of the heating profile remains to be 1.5 Mm. The total energy input for each of the line profiles shown in the right column is identical. The time for each line profile in the right column is marked with vertical dashed lines in the left column. }
\label{ib01all}
\end{figure}

\subsection{Parameter study}
\label{param}
Both EBs and UV bursts are believed to occur in the lower and partially ionized atmosphere, thus it would be difficult to infer how the energy is exactly deposited from observations. Inversions of line profiles, either through semi-empirical models \citep{2006fang,2014berlicki} or non-LTE inversion codes \citep{2019vissersb}, can deduce peaks in the temperature profiles. In order to reproduce such temperature profiles, we use a simple Gaussian function for the heating profile, with three free parameters (peak, width, and center). Judging from the above results, the heating center seems to be the most important parameter. Most optically thick lines form in a height range, and their behaviors can be quite different if different atmospheric layers are heated.   

It is, however, very interesting to explore how much influence the other two parameters (peak and width) would have on the line profiles. We choose EB2 and UVB1 as reference models, and vary these two parameters. We choose four values for each parameter, namely, 0.2, 0.5, 2, and 5 times the original value in the reference model, and show the results in Figs.~\ref{eb02all} and \ref{ib01all}. For all the line profiles shown in these figures, the total energy input is the same.

It is very clear that for both EBs and UV bursts, a larger heating peak would result in a more rapid rise in the line intensities (Figs.~\ref{eb02all}(a) and \ref{ib01all}(a)). Even with the same amount of input energy, a larger enhancement is expected for a larger heating peak (Figs.~\ref{eb02all}(b) and \ref{ib01all}(b)). However, the behavior of the spectral lines in response to the heating width is quite different. For the H$\alpha$ line in EBs, heating with a larger width also results in a more rapid rise in the line wing intensity (Fig.~\ref{eb02all}(c)), while after some time when the same amount of energy is deposited into the atmosphere, the line profiles do not vary much (Fig.~\ref{eb02all}(d)). For the \ion{Si}{4} 1403 \AA\ line in UV bursts, apart from the case with a very small heating width,  heating with a smaller width would normally results in a more rapid rise in the line center intensity (Fig.~\ref{ib01all}(c)). In this region, the main dissipation terms of the input energy are radiative losses and work done by pressure gradient. For the case with a smaller heating width, the amount of work done by pressure gradient at the height of the heating center (1.5 Mm) would be larger, while the optically thin radiative losses are smaller due to a lower electron density. After subtracting these dissipation terms, it turns out that the increase in the internal energy is generally larger for the case with a smaller heating width. As for the line profiles, only cases with smaller widths show clear enhancement of the \ion{Si}{4} line (Fig.~\ref{ib01all}(d)). This is because the \ion{Si}{4} line forms in a certain temperature range. For a larger heating width but a shorter heating time, one would expect no response of the \ion{Si}{4} line, since it  takes some time to heat the atmosphere until there are enough \ion{Si}{4} atoms.  When these two parameters are combined together, the influence of the heating peak dominates (Figs.~\ref{eb02all}(e)--(f) and \ref{ib01all}(e)--(f)).

The evolution of the line intensity shows different behavior in EBs and UV bursts, which is due to the difference in the formation height of the H$\alpha$ and the \ion{Si}{4} line. The H$\alpha$ line wing forms in the TMR where the input energy can be effectively radiated away through optically thick line photons, thus one sees a rise--plateau pattern in the lightcurve (Fig.~\ref{eb02all}, left column). The \ion{Si}{4} line, however, forms higher where the input energy cannot be radiated away effectively, thus the local temperature rises very quickly. The number of the \ion{Si}{4} atoms rises first, and then falls when the temperature is so high that they get ionized, thus one sees a rise--fall pattern in the lightcurve (Fig.~\ref{ib01all}, left column). For each case in the EB group or UV burst group, the pattern is similar and does not depend on the heating parameters, and one would expect similar line profiles at a certain time in each case, with a different total deposited energy. In our simulations, we choose a set of heating parameters that could show typical EB and UV burst features.

\subsection{The FUV continuum enhancement of EBs}
The enhancement of the FUV continuum is the most striking feature in our EB simulations. Previous observations confirmed that EBs are visible in the FUV continuum at 1600 \AA\ and 1700 \AA\ \citep{2013vissers,2015rezaei,2016tian,2017chen}. And it is very promising to detect EBs from these wavebands \citep{2019vissers}. However, there are rare reports of a significant enhancement near the 1400 \AA\ continuum. The observed  enhancement only appears in few cases and is  less than one order of magnitude \citep{2016tian,2019chen}.  Since most previous IRIS observations in the FUV waveband lack radiometric calibration, it would be difficult to directly compare the intensity values.

The FUV continua from 1400 \AA\ to 1700 \AA\ are mainly contributed by bound-free transitions of neutral metals \citep{2019simoes}, and thus their formation height should not vary much. In fact,   \cite{2016tian} showed that if the FUV continuum near 1400 \AA\ is enhanced, then the AIA 1700 \AA\ image is very likely to brighten up in the same region. 

The FUV continuum near 1400 \AA\ is formed in the lower chromosphere. Therefore,  if heating is constrained to the {deep} photosphere (EB1 and EB2 in the first few seconds), then the FUV continuum has little response to it (Fig.~\ref{si4cont}). \cite{2019seo} also calculated an EB model with heating  in the {deep} photosphere, and they were unable to reproduce enhancements in the FUV continua near 1600 \AA\ and 1700 \AA. These  spectral features shown in our simulations are supported by some observations, such as the EB (with no association of UV bursts) reported by \cite{2020ortiz} and the weak EB reported by \cite{2017honga}. 

However, there are still some observations of EBs that can not fit into the above category. An example is the strong EB reported by \cite{2017honga}, which shows clear emission peaks of the H$\alpha$ line wings and an enhancement in the AIA 1700 \AA\ emission, but no response in the IRIS 1400 \AA\ SJI. This lack of enhanced FUV emission in the IRIS SJI could be due to a low signal-to-noise ratio of the FUV band since the exposure time in this observation is too short. Some other events, which show response in the AIA 1600 and 1700 \AA\ continua but no response in the FUV continuum near the IRIS \ion{Si}{4} 1403 \AA\ line, are actually EBs associated with UV bursts. In these events, the chromosphere can be heated as a result of the UV burst, and the metal lines in the FUV waveband (e.g. the \ion{C}{4}, \ion{Si}{2}, and \ion{He}{2} lines), with formation temperatures of $\log T=4.2$--$5.0$ K \citep{2019simoes}, could also contribute to an enhancement in the integrated AIA 1600 and 1700 \AA\ emission.

The FUV continuum is also enhanced during flares, and there have been many observations \citep{2014tian,2015li,2016warren,2018tian,2020yu,2020zhou}. Recent RHD simulations of solar flares showed that the intensity of the FUV continuum can reach $3\times10^{-9}$ erg s$^{-1}$ cm$^{-2}$ Hz$^{-1}$ sr$^{-1}$ for case 5F9$\delta$4$E_c$20 (similar to a C-class flare) \citep{2019kerrb}, which is similar to our EB3. We speculate that the expected increase in FUV continuum could be detected by IRIS if the exposure time is not too short.

More observations and model simulations of EBs, especially in the FUV wavelengths, are thus needed to explore the causes for the discrepancy between observations and simulations. It would be helpful if these observed EBs are not related to UV bursts, so that the different spectral features of EBs and UV bursts would not superpose.

\subsection{The \ion{Mg}{2} lines of EBs}
Previous observations have reported plenty of EB events with line wing enhancement in the \ion{Mg}{2} triplet lines (including the \ion{Mg}{2} 2791 \AA\ line) \citep{2015vissers,2017honga,2019vissersb,2020ortiz}. From simulations, we confirm that the line wings of the \ion{Mg}{2} 2791 \AA\ line, especially the emission peaks appearing in some cases,  are formed in the heated region in the lower atmosphere (Fig.~\ref{mg2form}). \cite{2017honga} suggested a positive relationship between the line wing intensities of the H$\alpha$ line and the \ion{Mg}{2} triplet lines.  A further statistical study, similar to \cite{2019vissers}, is needed in order to investigate the potential of detecting EBs using the \ion{Mg}{2} triplet lines.

The response of the \ion{Mg}{2} k line  to EB heating has also been reported in some previous literature. However, we find that only the outer wings of \ion{Mg}{2} k are significantly enhanced while  the k$_2$ peaks show no response to EB heating. In fact, the k$_2$ peaks are only enhanced in  cases of UV bursts, which is confirmed by recent observations \citep{2020ortiz}. 

\subsection{The flow velocity and line asymmetries of EBs}
In our simulation, the mass flow velocity that is related to an EB is very small (below 2 km s$^{-1}$), and it is difficult to obtain a large velocity in the absence of magnetic fields. The flow velocities that are inferred from observed line profiles, either using the bisector method or a cloud model, have typical values of below 10 km s$^{-1}$ \citep{2008matsumotoa,2013yang,2014hong,2017libbrecht}. Recent non-LTE inversions recovered the bi-directional jets that exceed 20 km s$^{-1}$ \citep{2019vissersb}. MHD simulations of EBs usually show bidirectional flows from several to tens of km s$^{-1}$, and are more comparable to observations \citep{2007isobe,2009archontis,2016ni,2017hansteen,2019hansteen}.

Due to the weak mass flow in our simulations, the resulted H$\alpha$ line asymmetry is also very weak. The red asymmetry of the wing emission in the \ion{Mg}{2} 2791 \AA\ line is more obvious.  The  synthetic line profiles in our simulation of EBs only show a red asymmetry, while in observations, both red and blue asymmetries could be present \citep{1972bruzek,1983kitai,1997dara,2006fang,2006socas,2010hashimoto,2013vissers}. In addition, a moving overlying canopy might also influence the line asymmetry \citep{1972bruzek,1983kitai,2011watanabe,2013rutten,2019vissers}.

\subsection{The formation height of UV bursts}
The formation height of  UV bursts has undergone a long discussion. Recent studies suggest that UV bursts are formed in the mid to upper chromosphere \citep{2017hansteen,2019chen,2019hansteen}. In our simulations, the cases of UVB1 to UVB3  reproduce well spectral features that are typical of UV bursts. In particular, the synthetic \ion{Si}{4} and \ion{Mg}{2} k line profiles are very similar to the observed ones. We successfully reproduce the double-peaked \ion{Si}{4} line profiles that have been frequently observed in previous studies \citep{2014peter,2015vissers,2016tian,2017hansteen,2020ortiz}. The line width, however, is not comparable to observations, which is a modeling limitation and is discussed further in Section~\ref{limit}.

For the H$\alpha$ line, we confirm that the line center can be enhanced as shown in recent simulations \citep{2017hansteen}. We also find that in some cases (like UVB1 to UVB3), the line center is blueshifted. Similarly, \cite{2020ortiz} observed a blueshifted H$\alpha$ line center in  UV bursts, and they interpreted it as a result of an accompanying  surge. 

The case of UVB4 shows different spectral features from the other three cases. Yet UVB4 can still be classified as a UV burst due to the enhancement in the \ion{Si}{4} line \citep{2018young}. The specific features of redshifted H$\alpha$ and \ion{Si}{4} lines have also been shown in previous observations of UV bursts  \citep{2020ortiz}.

\subsection{The co-existence of EBs and UV bursts}
\begin{figure}
\epsscale{1.2}
\plotone{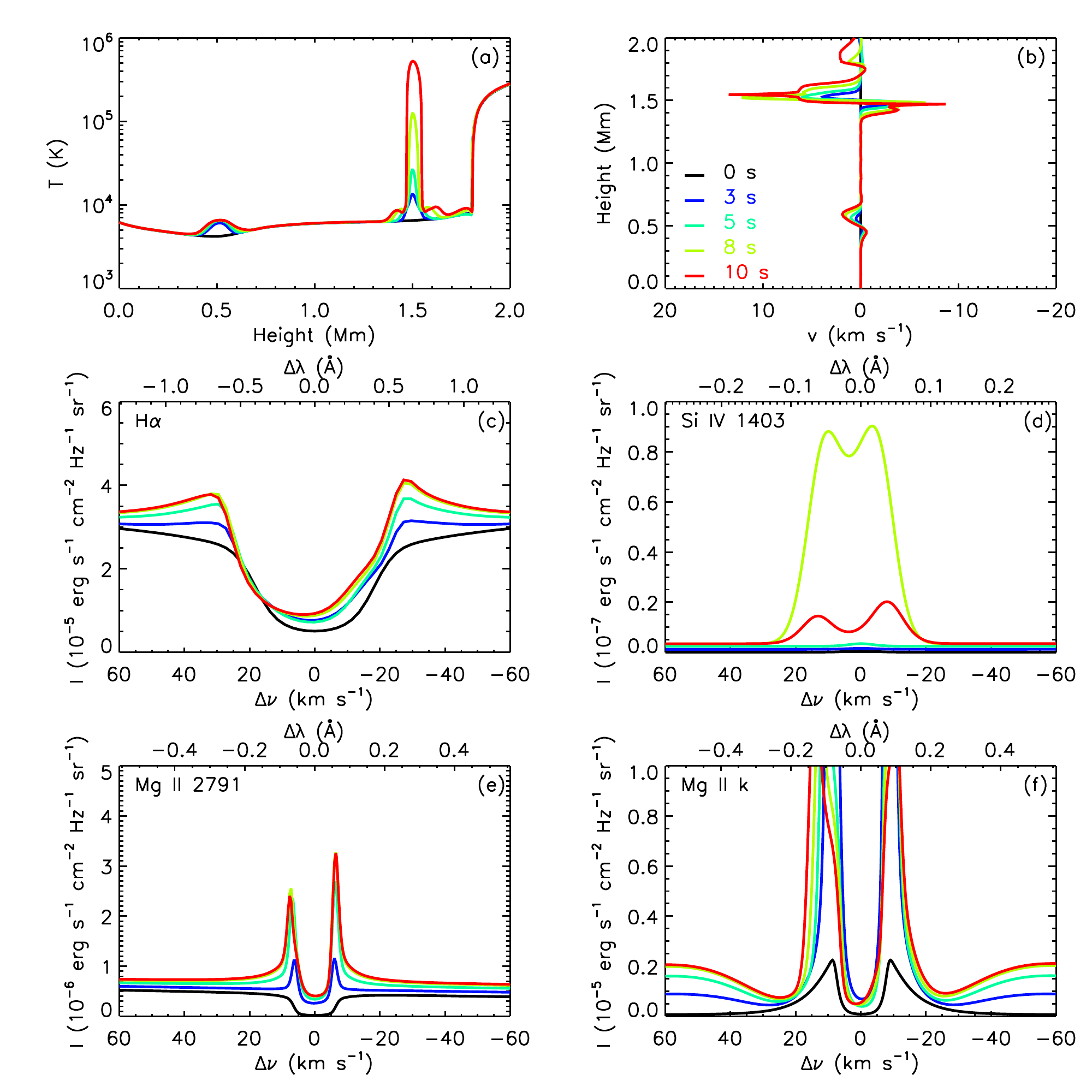}
\caption{Same as Fig.~\ref{eball}, but for a case with two heating sources. The heating profile is a combination of those of EB3 and UVB1. Colored lines denote different time. }
\label{ebib}
\end{figure}

EBs and UV bursts appear to be associated to each other in many events \citep{2015vissers,2016tian,2019chen,2020ortiz}, so it would be very interesting to explore posssible temperature structure that could generate both features simultaneously. Previous study of line inversions suggested two separate heating sources in the atmosphere \citep{2019vissersb}. Thus we run a new simulation with heating in both the TMR and the upper chromosphere. The new heating function is simply a combination of the heating functions in cases EB3 and UVB1, and we show the results in Fig.~\ref{ebib}. It is quite clear that since the two heating sources are quite far apart, they do not seem to influence each other. The typical EB features that are present in case EB3, as well as the typical UV burst features that are present in  case UVB1, both appear in the results here as expected (Fig.~\ref{ebib}(c)--(f)). Our results agree with previous observations \citep{2015vissers,2020ortiz} to a large extent. However, how these two heating sources are physically connected to each other would need further study with MHD simulations.

\begin{figure}
\epsscale{1.2}
\plotone{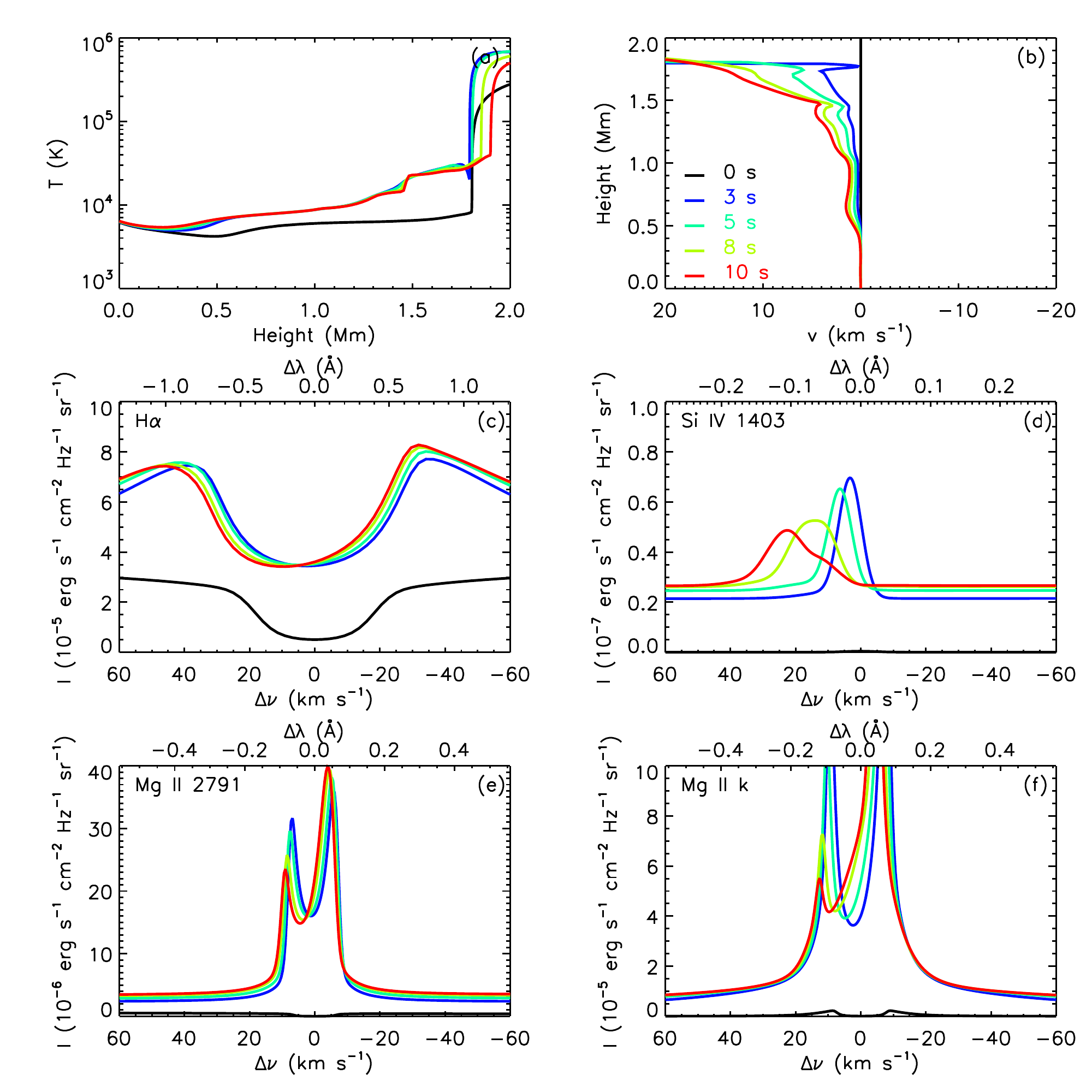}
\caption{Same as Fig.~\ref{eball}, but for a case with extended heating in the atmosphere. Colored lines denote different time.}
\label{ebib2}
\end{figure}

\begin{figure}
\epsscale{1.2}
\plotone{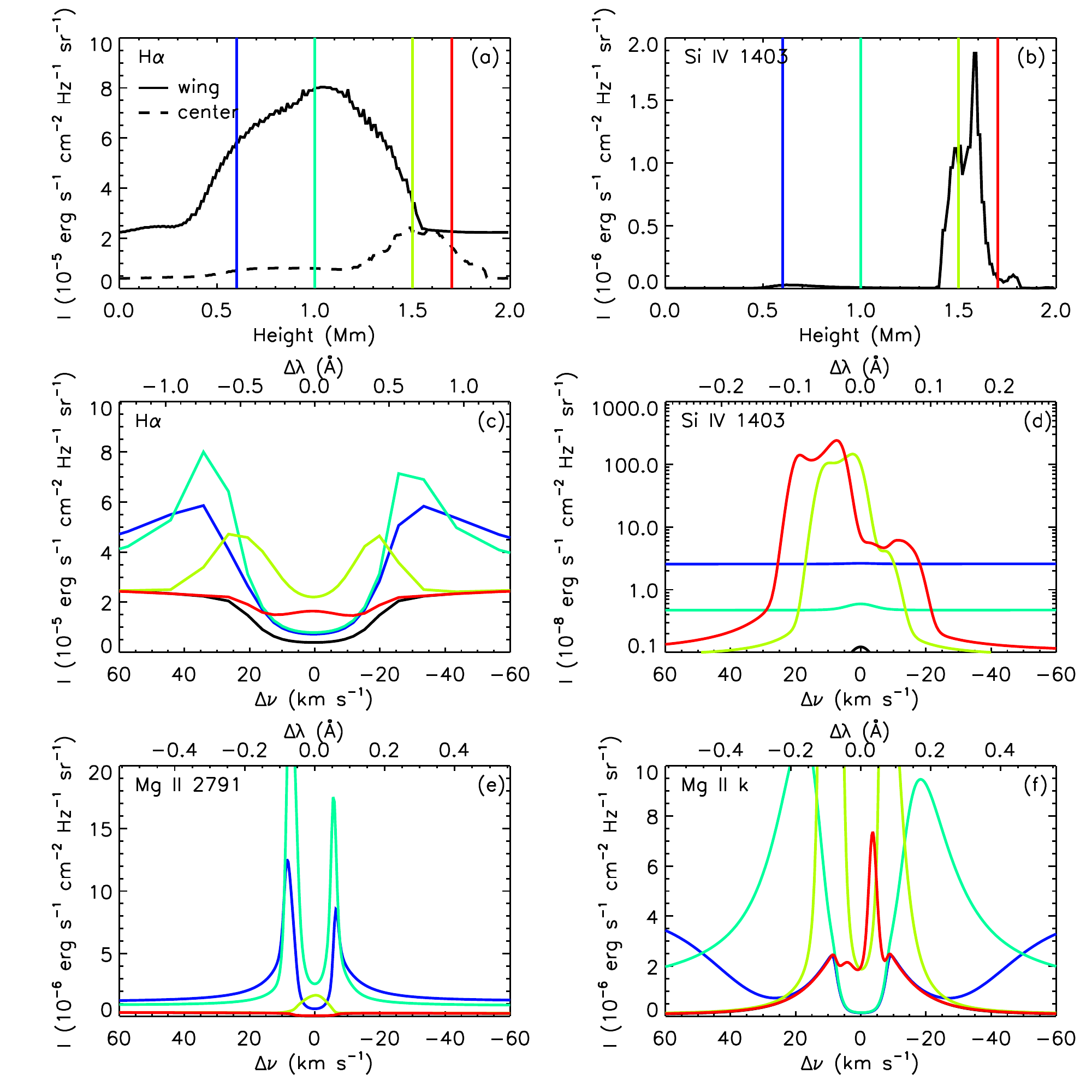}
\caption{Line intensities and profiles at different heights with a viewing angle $\mu=0.5$, for a case with extended heating in the atmosphere. (a) Line intensities of the H$\alpha$ line center (dashed) and blue wing ($-0.75$ \AA, solid) at different heights. (b) Line intensities of the \ion{Si}{4} 1403 \AA\ line center at different heights. (c)--(f) Line profiles at selected heights, which are marked with colored vertical lines in panels (a) and (b). Black lines denote quiet-Sun profiles.}
\label{ebib2mu}
\end{figure}

\cite{2019hansteen} showed that an extended current sheet is likely to generate both EB and UV burst in the atmosphere. To mimic such a current sheet, we employ a Gaussian heating profile with a very large width. As shown in Fig.~\ref{heatingrate}, the heating profile is centered at 600 km, with a peak of 500 erg cm$^{-3}$ s$^{-1}$ and a width of 500 km. The results are shown in Fig.~\ref{ebib2}. In this case, there are prevailing upflows in the lower atmosphere, and all the line centers are blueshifted. Compared with the above case, all the line profiles except \ion{Si}{4} have similar shapes but are dramatically enhanced. The H$\alpha$ line wing intensity increases by 120\%, and the line core intensity increases by 900\%. The line profile still presents an EB-like shape, with an emissive wing and an absorptive core, and we speculate that the overlying canopy, which is absent in our model, could reduce the line core intensity to some extent. The \ion{Si}{4} line is enhanced and broadened, but with a single peak since there is no downflow in the line formation region. 

However, in observations we are not always observing vertically, thus EBs and UV bursts can be some distances apart in our field of view \citep[e.g.][]{2019chen}. We assume that the size of EBs/UV bursts is 0.\arcsec 4, and bury the above column atmosphere with extended heating into a quiet-Sun background atmosphere. We then recalculate the line profiles with a viewing angle of $\mu=0.5$, and show the results in Fig.~\ref{ebib2mu}. When different parts of the column atmosphere are observed from aside, the line intensities can vary. The intensity of the H$\alpha$ line wing ($-0.75$ \AA) is enhanced within the height range of 0.5--1.5 Mm of the heated atmosphere. This extended brightening is quite similar to the ``flame morphology'' in previous studies \citep{2015vissers,2019chen,2019hansteen}. There are two height ranges where the intensity of the \ion{Si}{4} line center is increased. The one in the lower chromosphere (around 0.6 Mm) is contributed by the FUV continuum, while the other  in the upper chromosphere (around 1.5 Mm) is contributed by the \ion{Si}{4} line.  The line profiles also vary when the line of sight (LOS) reaches different parts of the heated atmosphere. At a lower height (0.6 Mm), one sees EB features that are similar to our simulation results of an isolated EB in Section~\ref{res}; when the LOS gradually moves to a larger height, UV burst features begin to appear, and at 1.5 Mm, there appear to be both features.

\subsection{Limitations}
\label{limit}
In our simulations, we are unable to reproduce very broad \ion{Mg}{2} and \ion{Si}{4} line profiles as in the observations. Such a discrepancy originates from the limitations of our 1D RHD modeling. Since we do not consider magnetic fields, the heating rate, as a result of magnetic reconnections in the lower atmosphere, is manually set as a simple Gaussian function, which might differ from the real situation. Thus the heating rate might be underestimated at certain heights, where one also underestimate the electron density that could contribute to a wider line wing \citep{2017rubio,2019zhu}. The non-thermal broadening due to mass flows and turbulence is also underestimated because of the lack of reconnection details. 

The 1D nature of our simulations implies that the intensities at line centers might be overestimated, since the 3D radiative transfer effect could smear out some portion of the emission \citep{2012leenaarts,2013leenaarts}. In addition, the atmosphere used in our simulations is an averaged atmosphere for the quiet Sun. In reality, due to the complexity of the magnetic fields in the active region, the atmospheric structure could also be very complicated.  For example, in our simulations, the enhancement of H$\alpha$ line center intensity in EB cases could be due to the absence of a fibrillar canopy in the chromosphere that can lead to a strong absorption \citep{2014hong,2017hansteen,2019vissersb}.

\section{Conclusion}
In this paper, we explore the responses of various spectral lines and continua towards heating in different atmospheric layers. Generally speaking, we successfully reproduce the spectral features of typical EBs and UV bursts respectively. Our main results, along with previous observations and simulations, are listed in Table~\ref{table2} and can be summarized as follows.

1. Heating in the lower atmosphere can generate typical EB features. The evolution of the H$\alpha$ line wing intensity shows a rise--plateau pattern. Enhancement in the FUV continuum near 1400 \AA\ is predicted if heating extends above the TMR. The FUV continuum intensity of EB3 can reach $3.4\times10^{-9}$ erg s$^{-1}$ cm$^{-2}$ Hz$^{-1}$ sr$^{-1}$, which is similar to the case of a C-class flare \citep{2019kerrb}.

2. Heating in the mid to upper chromosphere can generate typical UV burst features. 
The evolution of the \ion{Si}{4} 1403 \AA\ line center intensity shows a rise--fall pattern. At a certain time,
the \ion{Si}{4} 1403 \AA\ line is enhanced and very broad, with two emission peaks corresponding to the associated bidirectional mass flows. The H$\alpha$ line center is also blueshifted as a result of the upflow. 

3. Heating in the transition region would result in redshifted \ion{Si}{4} and H$\alpha$ lines.

4. Heating with two sources in the atmosphere, or an extended source, could reproduce both EB and UV burst features simultaneously.

Most of our calculated spectral features agree with previous observations and simulations with only a few exceptions. The predicted enhancement of FUV continuum near 1400 \AA\ in EB cases is rarely observed, although there are frequent reports of brightenings in AIA 1600 \AA\ and 1700 \AA\ images. We speculate that these EBs might {originate from the deep photosphere}, or the signal-to-noise ratio of the IRIS FUV observations is too low. It is also possible that in some reported EBs,  the AIA image 1600 \AA\ and 1700 \AA\ emissions could be influenced  by a related UV burst, where the chromospheric metal lines are enhanced and increase the integrated FUV emissions  in the waveband. In addition, our simulations show an enhancement of the H$\alpha$ line center during EBs and relatively narrow \ion{Mg}{2} and \ion{Si}{4} lines  that are incomparable to {most} observations. These  are due to the limitations of our 1D RHD simulations.

There have been observations of EBs in 1400 \AA, 1600 \AA, and 1700 \AA\ images \citep{2015vissers,2016tian,2017chen,2020ortiz}, yet we still expect more to statistically check the visibility in all these three FUV bands, with a particular combination of IRIS and AIA observations. The response of the H$\alpha$ line in UV bursts \citep{2017hansteen,2020ortiz} is also an interesting topic for future observations. In addition, joint observations including other chromospheric lines like the \ion{He}{1} D$_3$ and 10830 \AA\ lines  \citep{2017libbrecht}, are   useful to restrict the physical models. On the other hand,  3D (R)MHD modeling is needed to take into account the complexity of the atmosphere and magnetic field structure, which can include the 3D effects that are missed in the 1D models. Besides the realistic 3D simulations \citep{2017hansteen,2019hansteen}, it would also be helpful to perform some simplified 3D experiments with different pre-defined magnetic structures, in order to explore the formation mechanism of these events.

\acknowledgments
We are grateful to the referee for constructive comments. We thank Graham Kerr for providing the \ion{Si}{4} model atom. This work was supported by NSFC under grants 11903020, 11733003, 11873095, 11703012, and 11961131002, and NKBRSF under grant 2014CB744203. Y.L. is supported by the CAS Pioneer Talents Program for Young Scientists and XDA15052200, XDA15320301, and XDA15320103.

\clearpage

\end{document}